\documentclass[5p,twocolumn]{elsarticle}   

\usepackage[utf8]{inputenc}
\usepackage{amssymb}
\usepackage{mathtools}
\usepackage{graphicx}
\usepackage{graphicx,xcolor}
\usepackage{algorithm}
\usepackage{algpseudocode}
\usepackage{subfigure}
\usepackage{booktabs}
\usepackage[font=footnotesize]{caption}
\usepackage{empheq}
\usepackage{tcolorbox}
\usepackage{courier}
\usepackage{amsthm}
\usepackage{cases}
\usepackage{enumerate}
\usepackage{hyperref}
\journal{Systems \& Control Letters}

\newtheorem{theorem}{Theorem}[section]
\newtheorem{proposition}[theorem]{Proposition}
\newtheorem{lemma}[theorem]{Lemma}

\theoremstyle{definition}
\newtheorem{remark}[theorem]{Remark}

\newcommand{\longthmtitle}[1]{\mbox{}{\textit{(#1):}}}

\newcommand{\real}{\ensuremath{\mathbb{R}}}
\newcommand{\realpos}{\ensuremath{\mathbb{R}_{>0}}}
\newcommand{\realnonneg}{\ensuremath{\mathbb{R}_{\ge 0}}}

\newcommand{\intpos}{{\mathbb{N}}}

\newcommand{\map}[3]{#1: #2 \rightarrow #3}
\newcommand{\setdef}[2]{\{#1 \; | \; #2\}}

\newcommand{\Ac}{\mathcal{A}}

\newcommand{\Ec}{\mathcal{E}}

\newcommand{\Gc}{\mathcal{G}}

\newcommand{\Ic}{\mathcal{I}}
\newcommand{\Jc}{\mathcal{J}}
\newcommand{\Kc}{\mathcal{K}}
\newcommand{\Lc}{\mathcal{L}}

\newcommand{\Nc}{\mathcal{N}}

\newcommand{\Qc}{\mathcal{Q}}

\newcommand{\oprocendsymbol}{\hbox{$\bullet$}}
\newcommand{\oprocend}{\relax\ifmmode\else\unskip\hfill\fi\oprocendsymbol}

\newcommand{\ones}{\mathbf{1}}
\newcommand{\zeros}{\mathbf{0}}

\newcommand{\diag}{{\rm diag}}

\newcommand{\closure}{_{\operatorname{cl}}}
\newcommand{\threshold}{^{\operatorname{th}}}

\begin{document}
\begin{sloppypar}

\begin{frontmatter}

\title{Reinforcement Learning for Distributed Transient Frequency Control \\ with Stability and Safety Guarantees} 

\author[UCSD]{Zhenyi Yuan\corref{correspondingauthor}}\ead{z7yuan@ucsd.edu}    
\author[CUHK]{Changhong Zhao}\ead{chzhao@ie.cuhk.edu.hk}              
\author[UCSD]{Jorge Cort\'es}\ead{cortes@ucsd.edu}  
\cortext[correspondingauthor]{Corresponding author}

\address[UCSD]{Department of Mechanical and Aerospace Engineering, University of California, San Diego, La Jolla, CA 92093, USA}  
\address[CUHK]{Department of Information Engineering, The Chinese University of Hong Kong, New Territories, Hong Kong SAR}             

\begin{keyword}                           
Distributed Control; Power Systems; Reinforcement Learning; Performance Guarantees.              
\end{keyword}                             

\begin{abstract}                          
This paper proposes a reinforcement learning-based approach for optimal transient frequency control in power systems with stability and safety guarantees. Building on Lyapunov stability theory and safety-critical control, we derive sufficient conditions on the distributed controller design that ensure the stability and transient frequency safety of the closed-loop system. Our idea of distributed dynamic budget assignment makes these conditions less conservative than those in recent literature, so that they can impose less stringent restrictions on the search space of control policies. 
We construct neural network controllers that parameterize such control policies and use reinforcement learning  to train an optimal one. Simulations on the IEEE 39-bus network illustrate the guaranteed stability and safety properties of the controller along with its significantly improved optimality.
\end{abstract}

\end{frontmatter}

\section{Introduction}

With modern power systems shifting from high-inertia traditional generations to low-inertia renewable resources, it is increasingly important to design control mechanisms that allow to operate frequency around its nominal value. To tackle the frequency control problem, the appeal of learning methods  lies in the convenience of incorporating large amounts of data and accounting for optimality considerations in the controller design. This paper serves as a contribution to the growing body of work that seeks to leverage learning in the synthesis of efficient decision-making mechanisms in power systems that have rigorous guarantees on stability and performance.

\subsection*{Literature Review}
Transient stability of power systems refers to its ability to regain operating equilibrium after disturbances, while retaining the state within operational margins. The literature has investigated optimal frequency control design for improving transient stability, e.g., load-side control~\cite{CZ-UT-NL-SL:14,EM-CZ-SL:17}, droop coefficient design \cite{SSG-CZ-ED-YCC-SVD:18}, 
and proportional-derivative control~\cite{LG-CZ-SHL:18}, to mention a few. These methods either rely on designing optimal linear feedback controllers offline or solving optimization problems in real time to obtain optimal control policies. While these approaches ensure transient stability, they do not strictly guarantee transient safety, as the frequency may enter unsafe regions before convergence. To address this,~\cite{YZ-JC:19-auto} combined Lyapunov stability analysis and safety-control methods to ensure both stability and transient safety. This approach was further combined with model predictive control in~\cite{YZ-JC:20-tcns,YZ-JC:21-auto}
to minimize the control effort by enhancing the cooperation among the nodes, at the cost of a significantly heavier computational burden.

Recent research has employed data-driven methods to improve frequency control design  without the restriction for the controllers to be linear or the need to solve computationally complex optimization problems in real time. Reinforcement learning (RL) has emerged as an attractive method to learn such control policies offline, see e.g.,~\cite{XC-GQ-YT-SL-NL:22}. In general, stability and safety of the closed-loop system are not guaranteed without additional design constraints on the learned policies. This has resulted in a number of works that developed stability~\cite{WC-YJ-BZ:22,WC-BZ:21,WC-JL-BZ:22,YS-GQ-SL-AA-AW:22,YJ-WC-BZ-JC:22-ojcsys} or safety~\cite{DT-BZ:22,TLV-SM-RH-QH:21,TLV-SM-TY-RH-JT-QH:21} guaranteed RL approaches to learn optimal controllers for  frequency~\cite{WC-YJ-BZ:22,WC-BZ:21,DT-BZ:22,YJ-WC-BZ-JC:22-ojcsys} and voltage control~\cite{WC-JL-BZ:22,YS-GQ-SL-AA-AW:22,TLV-SM-RH-QH:21,TLV-SM-TY-RH-JT-QH:21} in power systems. With respect to previous work, the main novelty here is that we develop an RL-based approach that jointly  guarantees stability and transient frequency safety. To achieve this, we go beyond purely decentralized controller designs and leverage the distributed cooperation among agents, so that they can share the disturbance to be balanced and ensure system stability.

\subsection*{Statement of Contributions}
We study optimal transient frequency control in power systems with dynamics described by the swing equations. We formulate an optimization problem to identify control designs that minimize the frequency deviation from the equilibrium and the control cost over time while ensuring asymptotic stability and transient frequency safety in the presence of disturbances. Leveraging notions of Lyapunov stability and safety-critical control, we identify constraints on the distributed controller design whose satisfaction automatically guarantees that the closed-loop system remains stable and the  transient  frequency stays within the desired safety bounds. These constraints use \emph{budgets} to break down the requirement of collectively satisfying an inequality to ensure stability into individual stability conditions, one per bus, in a way that is distributed and allows additional design flexibility for certain buses while having others compensate for it. These constraints define the search space of distributed, stable, and safe control policies. We leverage them to enforce appropriate  structural constraints on neural networks so that the resulting parameterized controller belongs to the search space and can approximate with arbitrary accuracy any of its elements. Finally, we use a recurrent neural network (RNN)-based RL framework to learn the optimal parameters for these neural networks. Simulation results of the designed controllers on the IEEE 39-bus power system validate  their guaranteed stability and transient safety as well as their significantly improved optimality.

\section{Preliminaries}\label{sec-2}

We introduce here some notations and basic notions from the algebraic graph theory and the swing dynamics for power systems.

\subsection{Notations}
Throughout this paper, we use $\intpos$, $\real$, $\realnonneg$ and $\realpos$ to denote the set of natural, real, nonnegative and positive real numbers, respectively. For $a,b\in\intpos$, let $[a,b]_{\intpos}\triangleq\{x\in\intpos\ |\ a\leqslant x\leqslant b\}$. For $\mathcal{C} \subset \real^{n}$, $\partial\mathcal{C}$ denotes its boundary. For $A \in \real^{m\times n}$, $[A]_i$ and $[A]_{ij}$ represent its $i$-th row and $(i,j)$-th element, respectively. We denote by $A^{\dagger}$ its unique pseudo-inverse and by ${\rm Range}(A)$ its column space. $\ones_n$ and $\zeros_n$ in $\real^n$ are vectors of all ones and zeros, respectively. A continuous function $\alpha:\real \rightarrow \real$ is of (extended) class-$\mathcal{K}$ if it is strictly increasing and $\alpha(0)=0$. Finally, $\|\cdot\|_1$, $\|\cdot\|$ and $\|\cdot\|_\infty$ are respectively 1-norm, Euclidean norm and infinity norm.

\subsection{Graph theory}
Here we present some basic notions in graph theory~\cite{FB-JC-SM:08cor}. Let $\Gc = (\Ic,\Ec)$ be an undirected graph, where $\Ic = \{1,\dots,n\}$ is the node set and $\Ec = \{e_1,\dots,e_m\} \subseteq \Ic \times \Ic$ is the edge set. Two nodes are neighbors if there exists an edge linking them. We denote by $\Nc_i$ the set of neighbors of node $i$. A path is an ordered sequence of nodes such that any pair of consecutive nodes in the sequence is an edge of the graph. The graph $\Gc$ is connected if there exists a path between any pair of nodes. The adjacency matrix $\Ac$ is defined by $[\Ac]_{ij}>0$ if $i$ and $j$ are neighbors, 0 otherwise. The Laplacian matrix $\Lc$ is defined as $[\Lc]_{ij} = - [\Ac]_{ij}$ for $i \neq j$, and $[\Lc]_{ii} = \sum_{j=1,j\neq i}^n [\Ac]_{ij}$. The value $0$ is an eigenvalue of $\Lc$ with eigenvector $\ones_n$. This eigenvalue is simple if and only if the graph is connected. For each edge $e_k \in \Ec$ with nodes $i,j$, we assign an arbitrary orientation so that either $i$ or $j$ is the source of $e_k$ and the other node is the target of $e_k$. Then the incidence matrix $B = (d_{ik}) \in \real^{n\times m}$ of graph $\Gc$ is defined as
\begin{align*}
    d_{ik} = \begin{cases}1 & \text { if node } i \text { is the source of edge } e_{k} \\ -1 & \text { if node } i \text { is the target of edge } e_{k} \\ 0 & \text { otherwise. }\end{cases}
\end{align*}

\subsection{Power network dynamics}
The power network is modeled by a connected undirected graph $\Gc = (\Ic , \Ec )$, where $\Ic = \{1,\dots,n\}$ is the set of buses and $\Ec \subseteq \Ic \times \Ic$ is the set of transmission lines. We assume each bus represents an aggregate area consisting of loads and generators. For each bus $i \in  \Ic$, we use $\theta_i \in \real, \omega_i \in \real, p_i \in \real, u_i \in \real$ to represent its voltage angle, frequency deviation (from the nominal value), uncontrolled active power injection, and controlled active power injection, respectively. 
The frequency dynamics is described by the swing equations \cite{PK-NJB-MGL:94}:
\begin{align}\label{eq:swing}
  \dot{\theta}_{i}(t) &= \omega_{i}(t), \\
  M_{i} \dot{\omega}_{i}(t) &= \!-\! D_{i} \omega_{i}(t) \!-\! \sum_{j \in \mathcal{N}_i} b_{i j} \sin \left(\theta_{i}(t) \!-\! \theta_{j}(t)\right) \!+\! u_{i}(t) \!+\! p_{i}, \notag
\end{align}
for all $i \in \mathcal{I}$, where $M_i, D_i \in \realnonneg$ are the inertia and damping
coefficients of bus $i$, respectively, and $b_{ij} \in \realpos$ is the absolute value of susceptance of the transmission line connecting buses $i$ and $j$. For simplicity, we assume $M_i$, $D_i$ are all positive.

Define vectors $\theta \triangleq [\theta_1,\dots,\theta_n]^\top \in \real^n$, $\omega \triangleq [\omega_1,\dots,\omega_n]^\top \in \real^n$ and $p \triangleq [p_1,\dots,p_n]^\top \in \real^n$. 
Let $B \in \real^{n\times m}$ be the incidence matrix under an arbitrary graph orientation, and define the voltage angle difference vector
\begin{align}\label{eq:newvariable}
  \lambda(t) \triangleq B^\top \theta(t) \in \mathbb{R}^{m}.
\end{align}
Denote by $Y_b \in \real^{m\times m}$ the diagonal matrix with $[Y_b]_{k,k} = b_{ij}$, for
each edge $k = 1,2,\dots,m$ that links nodes $i$ and $j$, and define $M \triangleq \diag(M_{1}, M_{2}, \ldots, M_{n}) \in \real^{n\times n}$, $D \triangleq \diag(D_{1}, D_{2}, \ldots, D_{n}) \in \real^{n\times n}$. We rewrite the dynamics \eqref{eq:swing} in a compact form in terms of $\lambda(t)$ and $\omega(t)$
as
\begin{align}\label{eq:newswing}
\dot{\lambda}(t) &=B^\top \omega(t) \notag\\
M \dot{\omega}(t) &=-D \omega(t) - B Y_{b} \sin \lambda(t) + u(t) + p,
\end{align}
where $u(t) \triangleq [u_1(t),\dots,u_n(t)]^\top \in \real^n$, and $\sin \lambda(t) \in \real^m$ is the component-wise sine value of $\lambda(t)$. Note that the transformation \eqref{eq:newvariable} enforces $\lambda(0) \in {\rm Range}(B^\top)$. We refer to any $\lambda(0)$ satisfying this condition as an \emph{admissible} initial value. For convenience, we use $x(t) \triangleq (\lambda(t),\omega(t)) \in \real^{m+n}$ to denote the collection of all the state variables, and skip writing its dependence on $t$ if the context is clear.

Let $L \triangleq B Y_b B^\top$, and define $\omega^{\infty} \triangleq \frac{\sum_{i=1}^{n} p_{i}}{\sum_{i=1}^{n} D_{i}}$, $\tilde{p} \triangleq p-\omega^{\infty} D \mathbf{1}_{n}$. According to \cite{FD-MC-FB:13}, if
\begin{align}\label{eq:anglecondition}
  \left\|L^{\dagger} \tilde{p}\right\|_{\Ec, \infty}<1,
\end{align}
where $\|y\|_{\Ec, \infty} \triangleq \max _{(i, j) \in \Ec}\left|y_{i}-y_{j}\right|$, then there exists $\lambda^{\infty}\in \Gamma \!\triangleq\! \setdef{\lambda}{|\lambda_{k}| \!<\! \pi/2, \forall k=1,...,m}$ unique in
$\Gamma\closure \!\triangleq\! \setdef{\lambda}{|\lambda_{k}| \!\leq\! \pi/2, \forall k=1,...,m}$ such that
\begin{align*}
  \tilde{p}=B Y_{b} \sin \lambda^{\infty} \text { and } \lambda^{\infty} \in \operatorname{Range}(B^\top).
\end{align*}
Provided $\lambda(0) \in \operatorname{Range}(B^\top)$,~\eqref{eq:newswing} with $u \equiv 0$ has a unique equilibrium $(\lambda^{\infty},\omega^{\infty}\ones_n)$, which is asymptotically stable.

\section{Problem Formulation}\label{sec-3}

Consider a power network modeled as in Section~\ref{sec-2}. Under the condition \eqref{eq:anglecondition}, the unforced system admits a unique locally asymptotically stable equilibrium point $(\lambda^{\infty},\omega^{\infty}\ones_n)$. However, in the presence of disturbances, the transient frequency can enter unsafe regions before convergence to the equilibrium. This can be caused, for instance, by a sudden change in load or generation. If such frequency excursions exceed certain bounds, they might in fact lead to failure of loads or generators. To address this issue, we need to design  feedback controllers $\{u_i\}_{i \in \Ic}$ to ensure each nodal frequency $\omega_i$ 
stays within its safety bounds
$[\underline{\omega}_i,\overline{\omega}_i]$ during transients, while preserving the asymptotic stability of~\eqref{eq:newswing}. We also seek to minimize the frequency deviation from the equilibrium and the control cost integrated over time. This gives rise to
\begin{subequations}\label{optimization}
\begin{align}
\min_{u} & \int_{t=0}^{T} \left\{\gamma \|\omega(t) - \omega^\infty\ones_n\|^2  + \|u(t)\|^2\right\} dt \label{optimization:cost}\\
\text {s.t.} \ &\dot{\lambda} = B^\top \omega \label{optimization:dynamics-1}\\
&M \dot{\omega} = -D \omega - B Y_{b} \sin \lambda +u + p, \label{optimization:dynamics-2}\\
&\lim_{t \rightarrow \infty} (\lambda,\omega) = (\lambda^\infty,\omega^\infty \ones_n), \ \underline{\omega}_i \leq \omega_i \leq \overline{\omega}_i, \label{optimization:stability-safety}
\end{align}
\end{subequations}
where $T$ is the time horizon of interest and $\gamma$ is a coefficient balancing control cost and frequency deviation. Here $p = p^{\text{nom}} + \Delta p$, where $p^{\text{nom}}$ is the nominal power injection and $\Delta p$ accounts for the disturbance. We assume $\Delta p$ vanishes in finite time, and hence only affects the transient behavior. Note that we do not make any assumption about  a bound for the disturbance $\Delta p$, and consequently we do not impose saturation limits  in~\eqref{optimization} for~$u$. The term $\gamma \|\omega - \omega^\infty \ones_n\| ^2$ in the objective function penalizes deviation from the equilibrium frequency, and can be interpreted as a soft constraint to provide approximate transient safety. Instead, the safety bound in \eqref{optimization:stability-safety} is a hard constraint to strictly guarantee  transient safety by prohibiting the frequency nadir going outside the safe region. We also require the designed controllers to be distributed, in the sense that each bus can implement $u_i(x,p)$ using its local information and the information from its neighboring buses and incident transmission lines.

The infinite-dimensional and nonlinear nature of the optimization~\eqref{optimization} makes it hard to solve. Reinforcement learning (RL) is an attractive approach to such a problem by employing the data from system executions to train a policy that maps states to input actions. This results in a learned controller with optimized performance for the given data, but does not guarantee the stability and safety of the closed-loop system. Instead, model-based methods leverage knowledge of the dynamics to synthesize feedback controllers that render the system stable and safe, but have trouble dealing with the infinite-dimensional nature of the optimization. The advantages and limitations of RL and model-based approaches motivate us to combine them by identifying conditions on the controller design that ensure stability and safety (cf. Section~\ref{sec-4}) and incorporating these conditions in the RL policy search (cf. Section~\ref{sec-5}).

\section{Search Space of Control Policies}\label{sec-4}
Here we identify constraints on the control design that ensure  transient frequency safety and asymptotic stability. These constraints define later the search space of control~policies.

\subsection{Constraint ensuring frequency invariance}
We first turn our attention to the identification of conditions on the controller design that ensure the transient safety requirement, i.e., $\omega_i(t)$ staying in $[\underline\omega_i,\overline\omega_i]$ for all $i \in \Ic$ and all $t\geq 0$. For convenience,
let $\Qc_{i} \triangleq\left\{x \mid \underline{\omega}_{\mathrm{i}} \leq \omega_{\mathrm{i}} \leq \overline{\omega}_{\mathrm{i}} \right\}, \forall i \in \Ic$. To make this set forward invariant, one simply needs to ensure that the time-derivative of the frequency is negative when $\omega_i =  \overline\omega_i$, positive when  $\omega_i = \underline\omega_i$, and anything when $\omega_i \in (\underline\omega_i,\overline\omega_i)$. However, such specification may result in discontinuous controllers. Instead, we seek a specification that 
gradually kicks in as the frequency reaches certain thresholds,
while retaining the stability properties of~\eqref{eq:newswing} in the absence of input when the frequency is inside the thresholds. Meanwhile, a dead zone $[\underline\omega_{i}\threshold , \overline\omega_{i}\threshold]$ is introduced to avoid over-reaction of the controller to small frequency deviations. Built on this idea, the next result identifies a sufficient condition for a continuous controller design to ensure forward invariance of the frequency-safe set $\Qc_i$. 

\begin{lemma}\longthmtitle{Sufficient condition for frequency invariance \cite[Lemma 4.4]{YZ-JC:19-auto}}\label{lem:frequencyinvariance}
 Assume the solution of~\eqref{eq:newswing} exists and is unique for every admissible initial condition. For each $i\in\Ic$, let $\overline\omega_{i}\threshold,\
  \underline\omega_{i}\threshold\in\real$ be such that $\underline\omega_{i}<\underline\omega_{i}\threshold <
  \overline\omega_{i}\threshold<\overline\omega_{i}$ and $\overline\alpha_{i}(\cdot)$
  and $\underline\alpha_{i}(\cdot)$ be class-$\mathcal{K}$ functions.  If
  for all $x\in\real^{m+n}$ and~$p\in\real^{n}$,
    \begin{align}\label{eq:safety-constraint}
    \begin{cases}
     u_{i}(x,p) \leq \frac{\overline\alpha_{i}(\overline\omega_{i}-\omega_{i})}{\omega_{i}-\overline\omega_i\threshold} + q_{i}(x,p), & \omega_i > \overline\omega_{i}\threshold, \\
     u_{i}(x,p) \geq \frac{\underline\alpha_{i}(\underline\omega_{i}-\omega_{i})}{\underline\omega_{i}\threshold-\omega_{i}} + q_{i}(x,p), & \omega_i < \underline\omega_{i}\threshold,
    \end{cases}
    \end{align}
where $q_{i}(x, p) \triangleq D_{i} \omega_{i} + \left[B Y_{b}\right]_{i} \sin \lambda - p_{i}$, then $\Qc_i$ is a forward invariant set.
\end{lemma}

\subsection{Constraint ensuring asymptotic stability}\label{sec-4-1}

Here we derive a constraint on the control design that ensures asymptotic stability. We approach this by considering an energy function and restricting the input so that its time-derivative along the closed-loop dynamics is nonpositive.
Following~\cite{TLV-HDN-AM-JS-KT:18,YZ-JC:19-auto}, consider
\begin{align}\label{eq:energyfunction}
  V(\lambda, \omega) \triangleq \frac{1}{2} \sum_{i=1}^{n} M_{i}\left(\omega_{i}-\omega^{\infty}\right)^{2}+\sum_{j=1}^{m}\left[Y_{b}\right]_{j, j} a\left(\lambda_{j}\right),
\end{align}
where $a\left(\lambda_{j}\right) \triangleq \cos \lambda_{j}^{\infty} - \cos \lambda_{j} - \lambda_{j} \sin \lambda_{j}^{\infty} + \lambda_{j}^{\infty} \sin \lambda_{j}^{\infty}$. The derivative of $V$ along the dynamics~\eqref{eq:newswing} is given by
\begin{align}\label{eq:overall-stability-condition}
    \dot{V}(\lambda, \omega) \! = \!-\! \sum_{i=1}^{n} D_{i}\left(\omega_{i} \!-\! \omega^{\infty}\right)^{2} \!+\! \sum_{i=1}^n \left(\omega_{i} \!-\! \omega^{\infty}\right) u_{i}(x, p).
\end{align}
To ensure $\dot{V}(\lambda,\omega) \leq 0$, one can simply ask $u_i(x,p)$ to satisfy 
\begin{align}\label{eq:simple-stability-condition}
    - D_{i}(\omega_{i} - \omega^{\infty})^2 + (\omega_{i} - \omega^{\infty}) u_{i}(x, p) \leq 0,
\end{align}
for each $i \in \Ic$. This stability condition is convenient, from a network perspective, because it provides an individually decoupled constraint for each bus. This is essentially the approach taken in our previous work~\cite{YZ-JC:19-auto} and also in~\cite{WC-YJ-BZ:22}. Nevertheless, one can see that it is over-constraining, as the sum of all terms in~\eqref{eq:overall-stability-condition} is what needs to be nonpositive, not each individual summand. One could envision scenarios where some buses can deal with larger disturbances than others. In such cases, it would be advantageous to allow less capable buses to violate~\eqref{eq:simple-stability-condition} up to a level that can be compensated by more capable buses to still make the overall sum~\eqref{eq:overall-stability-condition} nonpositive.
Leveraging this insight, the next result generalizes the stability condition in~\cite[Lemma 4.1]{YZ-JC:19-auto}.

\begin{lemma}\longthmtitle{Sufficient condition for local asymptotic stability}\label{lem:stability}
  Consider system \eqref{eq:newswing} under condition \eqref{eq:anglecondition}. Further suppose that for every $i \in \Ic$, $u_{i}(x, p): \mathbb{R}^{m+n} \times \mathbb{R}^{n} \rightarrow \mathbb{R}$ is Lipschitz in~$x$. Let $c \triangleq \min _{\lambda \in \partial \Gamma\closure} V\left(\lambda, \omega^{\infty} \mathbf{1}_{n}\right)$
  and
  \begin{align}\label{Gamma}
    \Jc_\beta \triangleq\left\{(\lambda, \omega) \mid \lambda \in \Gamma\closure, V(\lambda, \omega) \leqslant c / \beta\right\}
  \end{align}
  with $\beta \in \realpos$. Suppose for every $i \in \Ic$, $x \in \real^{m+n}$, and $p \in \real^n$,
  \begin{align*}
  &
      \left(\omega_{i} \!-\! \omega^{\infty}\right) u_{i}(x, p) \!\leq\! \tilde{D}_i\left(\omega_{i} \!-\! \omega^{\infty}\right)^2 \!+\! b_i, && \text{if} \ \omega_{i} \neq \omega^{\infty}, 
\\
  &      u_{i}(x, p)=0, &&  \text{if} \ \omega_{i}=\omega^{\infty},
    \end{align*}
where $0 < \tilde{D}_i < D_i$ and $\sum_{i=1}^n b_i = 0$. Then, provided $\lambda(0) \in  {\rm Range}(B^\top)$ and $(\lambda(0),\omega(0)) \in  \Jc_\beta$ for some $\beta > 1$,
 \begin{enumerate}
    \item The solution of the closed-loop system exists and is unique for all $t \geq 0$;
    \item $\lambda(t) \in {\rm Range}(B^\top)$ and $(\lambda(t),\omega(t)) \in \Jc_\beta$ for all $t \geq 0$;
    \item $(\lambda^{\infty},\omega^{\infty}\ones_n)$ is stable, and $\lim_{t \rightarrow \infty} (\lambda(t),\omega(t)) = (\lambda^{\infty},\omega^{\infty}\ones_n)$.
  \end{enumerate}
\end{lemma}
\begin{proof}
Note that $\Jc_\beta$ is non-empty and compact. Hence if 2) holds, then 1) follows~\cite[Theorems 3.1 and 3.3]{HKK:02}. Therefore we focus on the statements 2)-3). From~\eqref{eq:overall-stability-condition},
\begin{align*}
 \dot{V}(\lambda, \omega)
&= -\sum_{i=1}^{n} D_{i}\left(\omega_{i} \!-\! \omega^{\infty}\right)^{2}+\sum_{i=1}^n \left(\omega_{i} \!-\! \omega^{\infty}\right) u_{i}(x, p) \\
&\leq -\sum_{i=1}^{n} (D_{i}-\tilde{D}_i) \left(\omega_{i} \!-\! \omega^{\infty}\right)^{2} \le 0 .
\end{align*}
Hence, given $(\lambda(0),\omega(0)) \in  \Jc_\beta$, we have $V(\lambda, \omega) \leq V(\lambda(0), \omega(0)) \leq c/\beta$, and 2) follows.
For 3), note that $V(\lambda, \omega) > 0$ for $(\lambda, \omega) \in \Jc_\beta \setminus (\lambda^{\infty},\omega^{\infty}\ones_n)$, and $V(\lambda^{\infty},\omega^{\infty}\ones_n) =0$, combined with $\dot{V}(\lambda, \omega) \leq 0$, implies that $(\lambda^{\infty},\omega^{\infty}\ones_n)$ is stable. Furthermore, noticing $\dot{V}(\lambda, \omega) = 0$ implies that $\omega = \omega^{\infty}\ones_n$, let $\Omega = \{(\lambda,\omega) \in \Jc_\beta \mid \omega = \omega^{\infty}\ones_n\}$, it is easy to see from \eqref{eq:newswing} that the largest invariant set in $\Omega$ is the point $\{(\lambda^\infty,\omega^{\infty}\ones_n)\}$. Then 3) follows the LaSalle Invariance Principle \cite[Theorem~4.4]{HKK:02}.
\end{proof}

The quantities $\{b_i\}_{i \in \Ic} \in \real^n$ in Lemma~\ref{lem:stability} correspond to the \emph{budgets} that allow some buses to violate the local condition~\eqref{eq:simple-stability-condition} 
and instead satisfy
\begin{align}\label{eq:stability-constraint}
      \begin{cases}
   u_{i}(x, p) \leq \tilde{D}_i(\omega_i-\omega^{\infty}) + \frac{b_i}{(\omega_i-\omega^{\infty})} & \omega_{i}>\omega^\infty, \\
   u_{i}(x, p) = 0 & \omega_{i} = \omega^\infty, \\
   u_{i}(x, p) \geq \tilde{D}_i(\omega_i-\omega^{\infty}) + \frac{b_i}{(\omega_i-\omega^{\infty})} & \omega_{i}<\omega^\infty,
   \end{cases} 
\end{align}
while ensuring system stability as long as $\sum_{i=1}^n b_i = 0$.
Notice that the introduction of the budgets may make the right-hand side of condition~\eqref{eq:stability-constraint} discontinuous, e.g., when $b_i \neq 0$ (the use of the LaSalle Invariance Principle requires $u_i$ to be 0 at $\omega^\infty$).
Note that the condition~\eqref{eq:stability-constraint} is more general than the stability condition in \cite[Lemma 4.1]{YZ-JC:19-auto}, which requires $u_i$ to have a different sign from $(\omega_i - \omega^\infty)$, and the stability condition in \cite[Theorem 1]{WC-YJ-BZ:22}, which further requires $u_i$ to be monotonically decreasing with $\omega_i - \omega^\infty$.

\subsection{Distributed dynamic budget assignment}
Before proceeding to the synthesis of distributed control policy, here we focus on the assignment of budgets introduced previously. Notice that both sufficient conditions \eqref{eq:safety-constraint} and \eqref{eq:stability-constraint} obtained above are naturally distributed, except for the non-sparse requirement $\sum_{i=1}^n b_i = 0$. Indeed, the satisfaction of this equality requires coordination across the buses. Interestingly, a static, a priori budget assignment in general does not work. This is because, if $b_i \neq 0$, then~\eqref{eq:stability-constraint} might require the control input to be infinitely large (instead of vanishing) when $\omega_i$ approaches $\omega^\infty$. Instead, 
the following result details a state-dependent budget assignment mechanism that ensures 
$b_i$ approaches zero as $\omega_i$ approaches~$\omega^\infty$ and enforces $\sum_{i=1}^n b_i = 0$  in a distributed way.

\begin{proposition}\longthmtitle{Distributed dynamic budget assignment}\label{prop:budget-assignment}
For $x \in \real^{m + n}$, let $\Ic\threshold \subseteq \Ic$ denote the set of buses satisfying $\omega_i \notin [\underline\omega_i\threshold,\overline\omega_i\threshold]$ and $\Ec\threshold \subseteq \Ec$ the set of edges between any pair of nodes in $\Ic\threshold$. Define the (possibly unconnected) state-dependent subgraph $\tilde{\Gc} = (\Ic,\Ec\threshold)$ of $\Gc = (\Ic,\Ec)$ and let $\tilde\Lc$ be its Laplacian matrix. Given 
$\xi \in \real^n$, let the budgets be assigned as $b_i = [\tilde{\Lc}]_i\xi$ for each $i \in \Ic$. Then the following holds:
\begin{enumerate}
\item $\sum_{i=1}^n b_i = 0$ always holds;
\item For each $i \in \Ic$, $b_i = 0$ whenever $\omega_i(t) \in [\underline{\omega}_i\threshold,\overline{\omega}_i\threshold]$.
\end{enumerate}
\end{proposition}
\begin{proof}
For 1), note that $\sum_{i=1}^n b_i = \sum_{i=1}^n [\tilde{\Lc}]_i\xi  = \ones_n^\top \tilde{\Lc} \xi$. Since 0 is an eigenvalue of $\tilde{\Lc}$ with eigenvector $\ones_n$, the conclusion follows. For 2),  according to the definition of $\tilde\Lc$, if $i \notin \Ic\threshold$, then $[\tilde{\Lc}]_i = \zeros_n^\top$, and hence $b_i=[\tilde{\Lc}]_i\xi = 0$.
\end{proof}

The underlying idea of Proposition~\ref{prop:budget-assignment} is that, instead of assigning $b_i$ directly to each bus $i \in \Ic$, we let each bus $i \in \Ic$ choose a value $\xi_i$ by itself and compute $b_i$ by exchanging information with its neighbors, utilizing the algebraic properties of Laplacian matrices to enforce $\sum_{i=1}^n b_i = 0$. For bus $i \in \Ic$, whenever $\omega_i(t) \in [\underline{\omega}_i\threshold,\overline{\omega}_i\threshold]$, the mechanism does not include it in the budget assignment process, and in that case its budget $b_i$ is simply zero. Finally, we remark that other dynamic budget assignment mechanisms, different from the one proposed here, might also work.

\subsection{Distributed, stable and safe control policies}\label{sec:all-reqs}
Here, we combine the results of the previous sections to identify the search space of distributed policies. In the next result, we propose a policy design to satisfy~\eqref{eq:safety-constraint} and \eqref{eq:stability-constraint}, which renders the closed-loop system stable and safe.

\begin{figure*}[ht]
\begin{align}\label{frequency-controller-constraint}
   \begin{cases}
   u_{i}(x, p) \leq \min \Big\{\tilde{D}_i(\omega_i-\omega^{\infty}) + \frac{b_i}{(\omega_i-\omega^{\infty})}, \frac{\overline{\alpha}_{i}(\overline{\omega}_{i}-\omega_{i})}{\omega_{i}-\overline{\omega}_{i}\threshold}+q_{i}(x, p)\Big\} & \omega_{i}>\overline{\omega}_{i}\threshold, \\
   u_{i}(x, p) = 0 & \underline{\omega}_{i}\threshold \leqslant \omega_{i} \leqslant \overline{\omega}_{i}\threshold, \\
   u_{i}(x, p) \geq \max \Big\{\tilde{D}_i(\omega_i-\omega^{\infty}) + \frac{b_i}{(\omega_i-\omega^{\infty})}, \frac{\underline{\alpha}_{i}(\underline{\omega}_{i}-\omega_{i})}{\underline{\omega}_{i}\threshold-\omega_{i}}+q_{i}(x, p)\Big\} & \omega_{i}<\underline{\omega}_{i}\threshold.
   \end{cases}
\end{align}
\hrulefill
\end{figure*}

\begin{figure}[htb]
    \centering
    \subfigure[$b_i >0 $]{
        \includegraphics[scale=0.39]{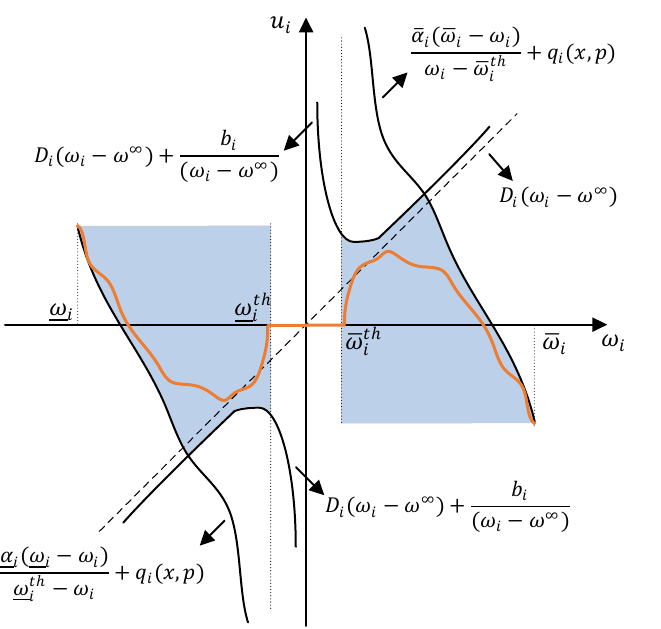}
    }
    \hspace{-.25in}
    \subfigure[$b_i <0 $]{
	\includegraphics[scale=0.39]{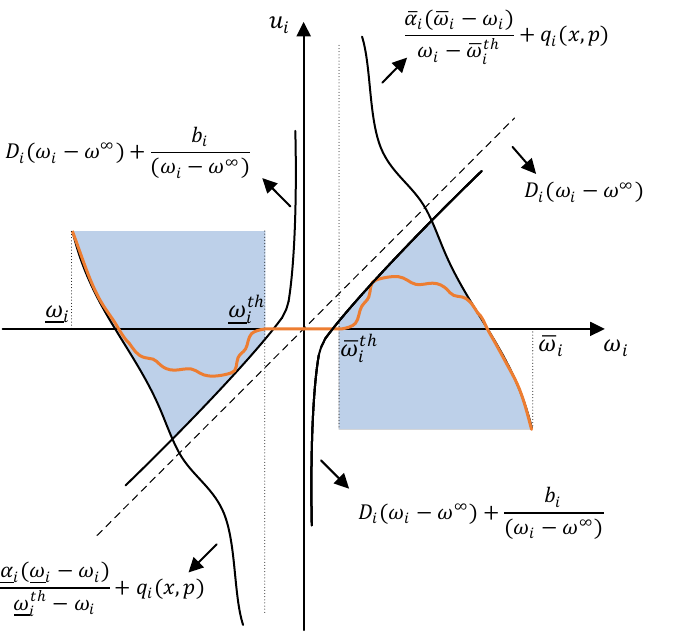}
    }
  \caption{The colored region shows the search space for the controllers satisfying by~\eqref{frequency-controller-constraint}, cf. Theorem~\ref{thm:frequency-controller}, which ensures asymptotic stability and transient safety. The orange curve is an instance of a controller in the specified search space. The sign of the budget captures whether bus $i$ (a) violates~\eqref{eq:simple-stability-condition} or (b) compensates it up to a certain amount to ensure the overall system stability.}\label{fig:search-space}
\end{figure}

\begin{theorem}\longthmtitle{Distributed control policies with asymptotic stability and transient safety guarantees}\label{thm:frequency-controller}
Given thresholds $\underline\omega_{i}\threshold$, $\overline\omega_{i}\threshold$ such that 
 $\omega^\infty \in (\underline{\omega}_{i}\threshold,\overline{\omega}_{i}\threshold)$. Let 
$\xi \in \real^n$ satisfy $\|[\Lc]_i\|_1 \|\xi\|_\infty \leq \min \{\tilde{D}_i (\overline\omega_i\threshold - \omega^\infty)^2, \tilde{D}_i( \underline\omega_i\threshold - \omega^\infty)^2\}$ for all $i \in \Ic$.
Under condition \eqref{eq:anglecondition}, consider the system~\eqref{eq:newswing} with a Lipschitz control policy satisfying~\eqref{frequency-controller-constraint} with budgets $b_i = [\tilde{\Lc}]_i\xi$ for each $i \in \Ic$, where $\tilde{\Lc}$ is defined in Proposition~\ref{prop:budget-assignment}. 
If $\lambda(0) \in  {\rm Range}(B^\top)$ and $(\lambda(0),\omega(0)) \in  \Jc_\beta$ for some $\beta > 1$, then the following holds:
\begin{enumerate}
    \item The solution of the closed-loop system exists and is unique for all $t \geq 0$;
    \item $\lambda(t) \in {\rm Range}(B^\top)$ and $(\lambda(t),\omega(t)) \in \Jc_\beta$ for all $t \geq 0$;
    \item $(\lambda^{\infty},\omega^{\infty}\ones_n)$ is stable, and $\lim_{t \rightarrow \infty} (\lambda(t),\omega(t)) = (\lambda^{\infty},\omega^{\infty}\ones_n)$;
    \item For each $i \in \Ic$, if $\omega_i(0) \in [\underline{\omega}_i,\overline{\omega}_i]$, then $\omega_i(t) \in [\underline{\omega}_i,\overline{\omega}_i]$ for all $t>0$;
    \item The budgets vanish in finite time, i.e., there exists a time $t_0 \!>\! 0$ such that $b_i \!=\! 0$ for all $t \!>\! t_0$ and all $i \!\in\! \Ic$.
 \end{enumerate}
\end{theorem}
\begin{proof}
 Statements 1)-4) readily follow Lemmas~\ref{lem:frequencyinvariance} and~\ref{lem:stability}  if \eqref{frequency-controller-constraint} (i) ensures that~\eqref{eq:safety-constraint} and \eqref{eq:stability-constraint} hold and (ii) defines a specification that can be satisfied by a Lipschitz controller. For (i), from Proposition~\ref{prop:budget-assignment} we have $\sum_{i=1}^n b_i = 0$, and therefore \eqref{frequency-controller-constraint} implies both \eqref{eq:safety-constraint} and \eqref{eq:stability-constraint}. For (ii), notice that the only problem is that \eqref{frequency-controller-constraint} requires $u_i = 0$ when $\omega_i \in [\underline\omega_i\threshold,\overline\omega_i\threshold]$. Hence, to guarantee it admits a Lipschitz controller, we need to show that $u_i$ can be chosen as $0$ right after $\omega_i$ passes the thresholds $\underline\omega_i\threshold$ and $\overline\omega_i\threshold$. Hence, it suffices to show
 that $\lim_{\omega_i \to (\overline\omega_i\threshold)^+} \min \{\tilde{D}_i(\omega_i-\omega^{\infty}) + \frac{[\tilde{\Lc}]_i\xi}{(\omega_i-\omega^{\infty})}, \frac{\overline{\alpha}_{i}(\overline{\omega}_{i}-\omega_{i})}{\omega_{i}-\overline{\omega}_{i}\threshold}+q_{i}(x, p)\} \geq 0$ and $\lim_{\omega_i \to (\underline\omega_i\threshold)^-} \max \{\tilde{D}_i(\omega_i-\omega^{\infty}) + \frac{[\tilde{\Lc}]_i\xi }{(\omega_i-\omega^{\infty})}, \frac{\underline{\alpha}_{i}(\underline{\omega}_{i}-\omega_{i})}{\underline{\omega}_{i}\threshold-\omega_{i}}+q_{i}(x, p)\} \leq 0$.
 Now we only show the case when $\omega_i \to (\overline\omega_i\threshold)^+$, since the other case can be proved similarly. Note that
$\lim_{\omega_i \to (\overline\omega_i\threshold)^+} \frac{\overline{\alpha}_{i}(\overline{\omega}_{i}-\omega_{i})}{\omega_{i}-\overline{\omega}_{i}\threshold}+q_{i}(x, p) = +\infty$, and hence the minimum is attained by the first term. We end the proof by noting that $\lim_{\omega_i \to (\overline\omega_i\threshold)^+}  \tilde{D}_i(\omega_i-\omega^{\infty}) + \frac{[\tilde{\Lc}]_i\xi}{(\omega_i-\omega^{\infty})} \geq \tilde{D}_i(\overline\omega_i\threshold-\omega^{\infty}) - \tilde{D}_i(\overline\omega_i\threshold-\omega^{\infty}) = 0$, where we have employed the fact that $|[\tilde\Lc]_i\xi| \leq \|[\Lc]_i\|_1 \|\xi\|_\infty$.
Finally, for 5), the asymptotic convergence established in 3) indicates that there exists a $t_0>0$ such that $\underline \omega_i\threshold \leq \omega_i(t) \leq  \overline \omega_i\threshold$ holds for all $t > t_0$ and all $i \in \Ic$. Together with Proposition~\ref{prop:budget-assignment}, the result follows. 
\end{proof}

Theorem~\ref{thm:frequency-controller} provides a characterization of the search space of distributed control policies that guarantee asymptotically stable and transient-safe closed-loop systems. Fig.~\ref{fig:search-space} illustrates the search space.

\section{Synthesis of Distributed Neural Network Controllers}\label{sec-5}
In this section, we construct neural networks that parameterize control policies satisfying the requirements in Section~\ref{sec:all-reqs}, 
and then apply an RNN-based RL framework to train an optimal one.

\subsection{Selecting  class-$\Kc$ functions and frequency thresholds}

The condition~\eqref{frequency-controller-constraint} obtained in Theorem \ref{thm:frequency-controller} depends upon the class-$\Kc$ functions $\overline{\alpha}_i, \underline{\alpha}_i$ and the frequency thresholds $\underline{\omega}_{i}\threshold,\overline{\omega}_{i}\threshold$.
Their choice affects the search space of control policies.
One can make specific choices for these design parameters according to practical considerations. Alternatively, one can use neural networks to parameterize and train them along with the control policy. Note that parameterizing $\overline{\omega}\threshold_i$ and $\underline{\omega}\threshold_i$ is easy since they are static values instead of functions. The more difficult task of parameterizing $\overline{\alpha}_i$ and $\underline{\alpha}_i$ requires the neural networks to be strictly monotone. Approaches to this include structure-based \cite{HD-MV:10} and verification-based \cite{XL-XH-NZ-QL:20} methods.
Here, we adopt the single hidden layer monotone neural network design in \cite{WC-YJ-BZ:22}, which achieves universal approximation, yet is easy to implement. We provide next the details of the proposed parameterization method.

\begin{lemma}\longthmtitle{Neural network parameterization of class-$\Kc$ functions}\label{lem:NN-kappa}
Let $\sigma(x) = \max(0,x)$ be the ReLU function. For each $i\in\Ic$, let
\begin{align*}
    \overline{\alpha}_i(\omega_i) = \overline{z}_i^+ \sigma(\ones_m \omega_i + \overline{c}_i^+) + \overline{z}_i^- \sigma(- \ones_m \omega_i + \overline{c}_i^-),
\end{align*}
where $\overline{c}_i^+,\overline{c}_i^- \in \real^m$ are bias vectors with $m$ hidden units satisfying $[\overline{c}_i^+]_1 = 0$, $[\overline{c}_i^+]_j \leq [\overline{c}_i^+]_{j-1}$ (resp. $[\overline{c}_i^-]_1 = 0$, $[\overline{c}_i^-]_j \leq [\overline{c}_i^-]_{j-1}$) for $j \in [2,m]_\intpos$, and $\overline{z}_i^+,\overline{z}_i^- \in \real^{1\times m}$ are weight vectors satisfying $\sum_{j=1}^{\ell} [\overline{z}_i^+]_{1,j} > 0$ (reps. $\sum_{j=1}^{\ell} [\overline{z}_i^-]_{1,j} < 0$) for $\ell \in [1,m]_\intpos$. Then, $\overline{\alpha}_i$ is of class-$\Kc$. Furthermore, for any class-$\Kc$ function $\kappa$ and given any compact domain~$K \subset \real$ and  $\epsilon > 0$, there exist $\overline{z}_i^+,\overline{z}_i^-,\overline{c}_i^+,\overline{c}_i^-$ and $m$ such that $|\kappa(\omega_i) - \overline{\alpha}_i(\omega_i)| < \epsilon$ for all $\omega_i \in K$.
\end{lemma}

We omit the proof of this result, but note that it is analogous to the proof of \cite[Theorem~2]{WC-YJ-BZ:22}. The underlying idea is to construct a piece-wise linear approximation of a nonlinear function in which every linear segment is strictly increasing. An arbitrary accuracy of the approximation can be achieved given sufficiently many neurons.
Also note that $\underline{\alpha}_i$ can be constructed in the same way with weight vectors $\underline{z}_i^+,\underline{z}_i^-$ and bias vectors $\underline{c}_i^+,\underline{c}_i^-$.

\begin{lemma}\longthmtitle{Neural network parameterization of frequency threshold}\label{lem:NN-threshold}
Let $\varsigma(x) = \frac{1}{1+e^{-x}}$ be the sigmoid function. For each $i \in \Ic$, let 
\begin{align*}
    \underline{\omega}_{i}\threshold = (\omega^\infty - \underline{\omega}_i) \varsigma(v_i^+) + \underline{\omega}_i, \ \overline{\omega}_{i}\threshold = (\omega^\infty - \overline{\omega}_i) \varsigma(v_i^-) + \overline{\omega}_i,
\end{align*} 
where $v_i^+, v_i^- \in \real$ are biases. Then $\underline{\omega}_{i}\threshold$ and $\overline{\omega}_{i}\threshold$ approximate any values in $(\underline{\omega}_i,\omega^\infty)$, and $(\omega^\infty,\overline{\omega}_i)$, respectively.
\end{lemma}

The proof of Lemma \ref{lem:NN-threshold} readily follows the definition of the sigmoid function. 

\subsection{Neural network controller design}

We first give the final ingredient to parameterize control policies that satisfy condition~\eqref{frequency-controller-constraint} using neural networks.
The next result provides a parameterization of
any function $\omega_i \mapsto f_i(\omega_i)$ satisfying $f_i(\omega_i)=0$ for $\omega_i \in [\underline{\omega}_{i}^{\mathrm{th}},\overline{\omega}_{i}^{\mathrm{th}}]$. 

\begin{lemma}\longthmtitle{Neural network parameterization of $f_i$}\label{lem:NN-control-wo-constraint}
For each $i \in \Ic$, let
\begin{align*}
    f_i(\omega_i) \!=\!  q_i^+ \sigma(\ones_m (\omega_i \!-\! \overline{\omega}_{i}\threshold) \!+\! r_i^+) \!+\! q_i^- \sigma(\!-\! \ones_m (\omega_i \!-\! \underline{\omega}_{i}\threshold) \!+\! r_i^-),
\end{align*}
where $r_i^+, r_i^- \in \real^m$ are bias vectors with $m$  hidden units satisfying $[r_i^+]_j \leq 0$ and $[r_i^-]_j \leq 0$ for all $j \in [1,m]_\intpos$, and $q_i^+, q_i^- \in \real^{1\times m}$ are weight vectors. Then, $f_i(\omega_i)=0$ for $\omega_i \in [\underline{\omega}_{i}\threshold,\overline{\omega}_{i}\threshold]$. Moreover, for any Lipschitz function $\map{g_i}{\real}{\real}$  satisfying $g_i(\omega_i)=0$ for $\omega_i \in [\underline{\omega}_{i}\threshold,\overline{\omega}_{i}\threshold]$ and given any compact domain $K \subset \real$ and $\epsilon > 0$, there exists $q_i^+, q_i^-,r_i^+, r_i^-$ and $m$ such that $|f_i(\omega_i) - g_i(\omega_i)| < \epsilon$ for all~$\omega_i \in K$.
\end{lemma}

The proof of this result uses the definition of ReLU function and exploits a piece-wise linear approximation similar to that in Lemma \ref{lem:NN-kappa}.
Let $z := \{\overline{z}_i^+,\overline{z}_i^-,\underline{z}_i^+,\underline{z}_i^-\}_{i \in \Ic}$, $c := \{\overline{c}_i^+,\overline{c}_i^-,\underline{c}_i^+,\underline{c}_i^-\}_{i \in \Ic}$, $v := \{v_i^+,v_i^-\}_{i \in \Ic}$, $q := \{q_i^+,q_i^-\}_{i \in \Ic}$, $r := \{r_i^+,r_i^-\}_{i \in \Ic}$ and denote $\phi = \{z,c,v,q,r,\xi\}$. The following result constructs the distributed neural network controllers.

\begin{theorem}\longthmtitle{Distributed neural network controllers}\label{Thm:NN-control-policy}
For each $i \in \Ic$, let $\overline{\alpha}_i$, $\underline{\alpha}_i$, $\overline{\omega}_{i}\threshold$, $\underline{\omega}_{i}\threshold$, and $f_i$ be constructed according to Lemmas \ref{lem:NN-kappa}, \ref{lem:NN-threshold} and \ref{lem:NN-control-wo-constraint}, respectively.
Under the assumptions of Theorem~\ref{thm:frequency-controller}, 
let  $\overline{u}_{i,\phi}(x,p) = \min \{\tilde{D}_i(\omega_i-\omega^{\infty}) + \frac{[\tilde{\Lc}]_i \xi}{(\omega_i-\omega^{\infty})}, \frac{\overline{\alpha}_{i}(\overline{\omega}_{i}-\omega_{i})}{\omega_{i}-\overline{\omega}_{i}\threshold}+q_{i}(x, p) \}$ and $\underline{u}_{i,\phi}(x,p) = \max \{\tilde{D}_i(\omega_i-\omega^{\infty}) + \frac{[\tilde{\Lc}]_i \xi}{(\omega_i-\omega^{\infty})}, \frac{\underline{\alpha}_{i}(\underline{\omega}_{i}-\omega_{i})}{\underline{\omega}_{i}\threshold-\omega_{i}}+q_{i}(x, p)\}$. Then,
\begin{align}\label{eq:NN-control-policy}
    & u_{i,\phi}(x,p) \!=\! \frac{\sigma(\omega_i \!-\! \overline{\omega}_{i}\threshold)}{\omega_i \!-\! \overline{\omega}_{i}\threshold} \left[f_i(\omega_i) \!-\! \sigma(f_i(\omega_i) \!-\! \overline{u}_{i,\phi}(x,p))\right]  \notag \\ 
    & \qquad + \frac{\sigma(\underline{\omega}_{i}\threshold
     \!-\! \omega_i)}{\underline{\omega}_{i}\threshold
     \!-\! \omega_i} \left[f_i(\omega_i) \!+\! \sigma(\underline{u}_{i,\phi}(x,p) \!-\! f_i(\omega_i))\right] ,
 \end{align}
is a distributed control policy satisfying~\eqref{frequency-controller-constraint}.
Furthermore, any Lipschitz  control policy satisfying~\eqref{frequency-controller-constraint} can be approximated arbitrarily close by~\eqref{eq:NN-control-policy}.
\end{theorem}

The proof readily follows the universal approximation results in Lemmas \ref{lem:NN-kappa}, \ref{lem:NN-threshold} and \ref{lem:NN-control-wo-constraint}.

\subsection{Learning optimal control policy using RNN}

Having parameterized in Theorem~\ref{Thm:NN-control-policy} the search space identified in Section~\ref{sec:all-reqs}, here we describe an approach to train an optimal control policy adopting the RNN-based RL framework proposed in~\cite{WC-YJ-BZ:22}. To simulate the trajectories for training the neural network controller \eqref{eq:NN-control-policy}, we use a first-order Euler discretization with stepsize $\Delta t$ for problem \eqref{optimization}. Let $K$ be the total number of timesteps. The discrete-time optimization problem is
\begin{subequations}\label{eq:NN-optimization}
\begin{align}
\min_{\phi} & \frac{1}{K} \sum_{k=0}^{K-1} \gamma \|\omega(k) - \omega^\infty\ones_n\|^2 + \|u_{\phi}(k)\|^2 \label{eq:NN-optimization:cost}\\
\text {s.t.} \ &\lambda(k) = \lambda(k-1) + B^\top \omega(k-1) \Delta t \label{eq:NN-optimization:dynamics-1}\\
&M (\omega(k) - \omega(k-1)) =   \big[- D \omega(k-1)  \notag\\
& \quad - B Y_{b} \sin \lambda(k-1) + u_{\phi}(k-1) + p \big] \Delta t. \label{eq:NN-optimization:dynamics-2}
\end{align}  
\end{subequations}
where $u_{\phi} = [u_{1,\phi},\dots,u_{n,\phi}]^\top$. The learning algorithm works as follows. At the beginning of the training process, all parameters in $\phi$ are randomly generated. Training is implemented in a batch updating style, where the initial states $\omega(0)$ and $\lambda(0)$ in each batch are randomly generated.
In each episode, we use the current control policy $u_{\phi}$ to generate state trajectories of length $K$ for all batches through dynamics \eqref{eq:NN-optimization:dynamics-1},\eqref{eq:NN-optimization:dynamics-2}, and compute the loss function \eqref{eq:NN-optimization:cost}. The trainable parameters $\phi$ are updated by gradient descent on the loss function \eqref{eq:NN-optimization:cost} and converge to a local optimum. Note that the training could be done in a distributed fashion, see e.g.,~\cite{AN-AO-PAP:10,PS-JC:21-tcns} and references therein, since the cost~\eqref{eq:NN-optimization:cost} is separable and the controller~\eqref{eq:NN-control-policy} is distributed. The whole training process of the proposed reinforcement learning with budget (abbreviated \emph{RLb}) algorithm is described in Algorithm~\ref{alg:RL}.

\begin{algorithm}[ht]
  \caption{Training Process of RLb Algorithm}\label{alg:RL}
  \begin{algorithmic}[1]
  \Require ~Total time stages $K$, number of episodes $I$, batch size $H$, learning rate $\alpha$, parameters in~\eqref{eq:NN-optimization}
  \renewcommand{\algorithmicrequire}{\textbf{Input:}}
  \Require ~Initial neural network parameters $\phi$
  \For{episode $i = 1$ to $I$}
  \State ~Randomly generate initial states $\lambda^h(0)$, $\omega^h(0)$ for $h$-th batch, $h=1,...,H$
  \State ~Reset the state of cells in $h$-th batch as $\{\lambda^h(0),\omega^h(0)\}$
  \State ~RNN cells compute through $K$ stages to obtain the output $\gamma \|\omega(k) - \omega^\infty\ones_n\|^2  + \|u_\phi(k)\|^2$ for $k=1,...,K$
  \State ~Calculate the average loss of all the batches $\textsf{Loss} = \frac{1}{H}\sum_{h=1}^H \sum_{k=1}^{K} \gamma \|\omega^h(k) - \omega^\infty\ones_n\|^2  + \|u_\phi^h(k)\|^2$
  \State ~Update neural network parameters by passing \textsf{Loss} to Adam optimizer: $\phi \gets \phi - \alpha \text{Adam}(\textsf{Loss})$
  \EndFor
  \renewcommand{\algorithmicensure}{\textbf{Output:}}
  \Ensure ~Trained controllers $u_\phi$
\end{algorithmic}
\end{algorithm}

\begin{remark}\longthmtitle{Robust policy learning through Lipschitz regularization}
One way to enhance the robustness of the learned control policy is to add an additional regularization term in the cost function to promote the control policy by a small Lipschitz constant \cite{PP-AK-JB-PK-FA:21}. Here we implement this Lipschitz-regularized learning by adding an additional term $\rho \frac{1}{K-1} \sum_{k=1}^{K-1} \|u_{\phi}(k) - u_{\phi}(k-1)\|^2$ to \eqref{eq:NN-optimization:cost}, where the parameter $\rho$ controls the trade-off between robustness and optimality of the learned control policy. We implement this reguralization in the simulations below and illustrate its added robustness against state measurement noise. \oprocend
\end{remark}

\begin{figure}[t]
    \centering
    \includegraphics[width=.75\linewidth]{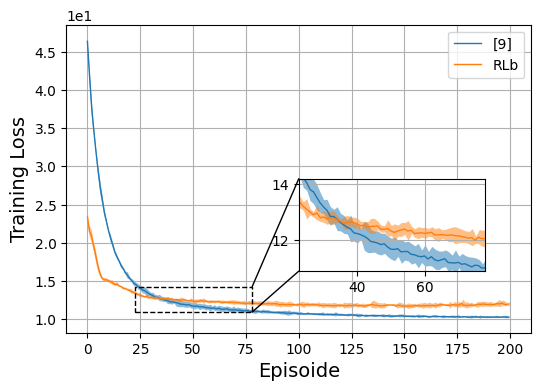}
    \caption{Comparison of average training loss curves and shaded error bars (representing the standard deviations) between the RL-based method in~\cite{WC-YJ-BZ:22} and the proposed RLb method based on 5 experiments. The latter has a warmer training start. Each method solves a different discrete-time optimization problem, which explains the convergence to different optimal values.}
    \label{fig:loss}
\end{figure}
\begin{figure*}[ht]
  \centering
   {\includegraphics[width=.9\linewidth]{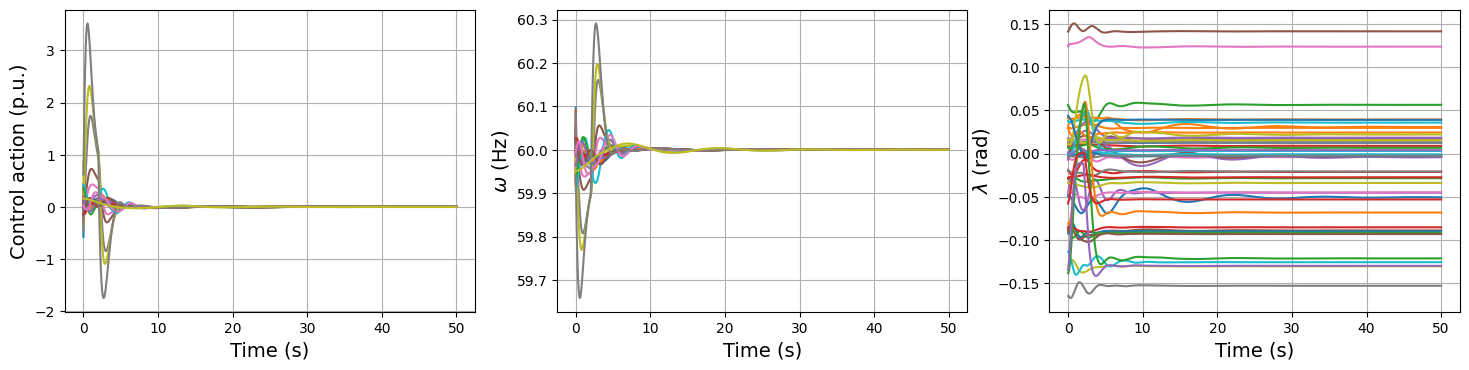}}\\
   {\includegraphics[width=.9\linewidth]{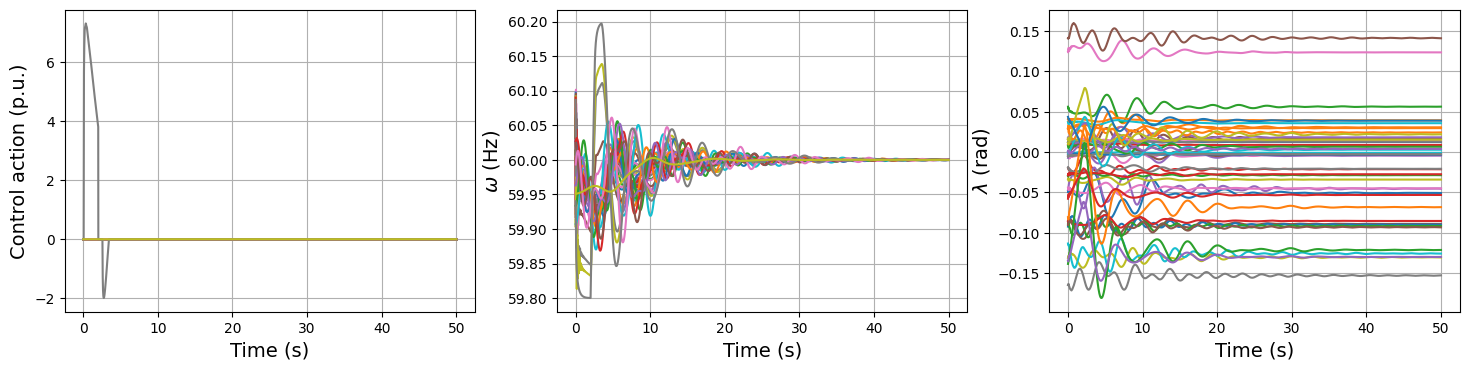}}\\
   {\includegraphics[width=.9\linewidth]{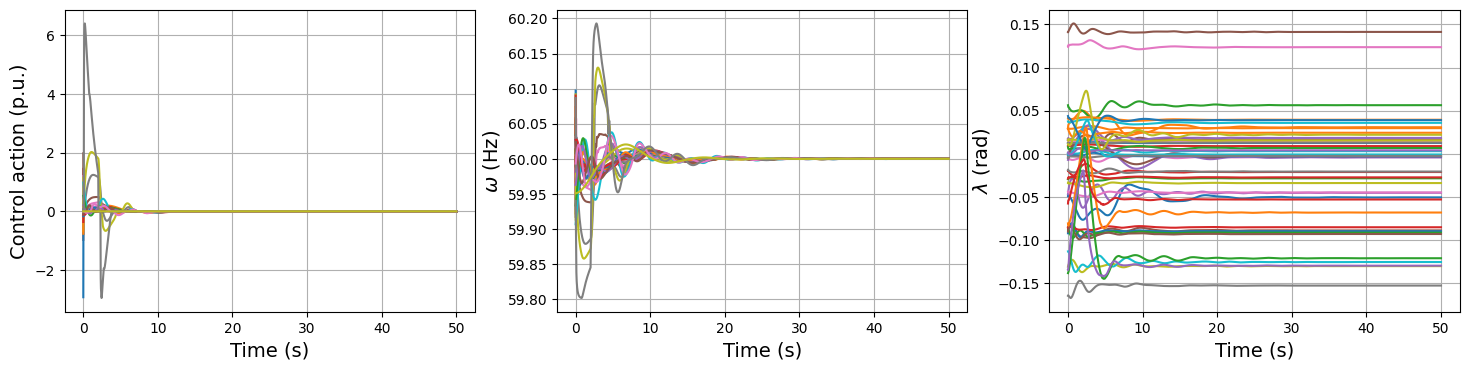}}\\
    \caption{Dynamics of the IEEE 39-bus network under the trained controllers via the RL-based method in~\cite{WC-YJ-BZ:22} (top), the controllers in~\cite{YZ-JC:19-auto} (middle), and the trained controllers via the RLb method in this paper (bottom).
 }\label{fig:sim}
\end{figure*}

\begin{figure*}[htb]
  \centering
{\includegraphics[width=.9\linewidth]{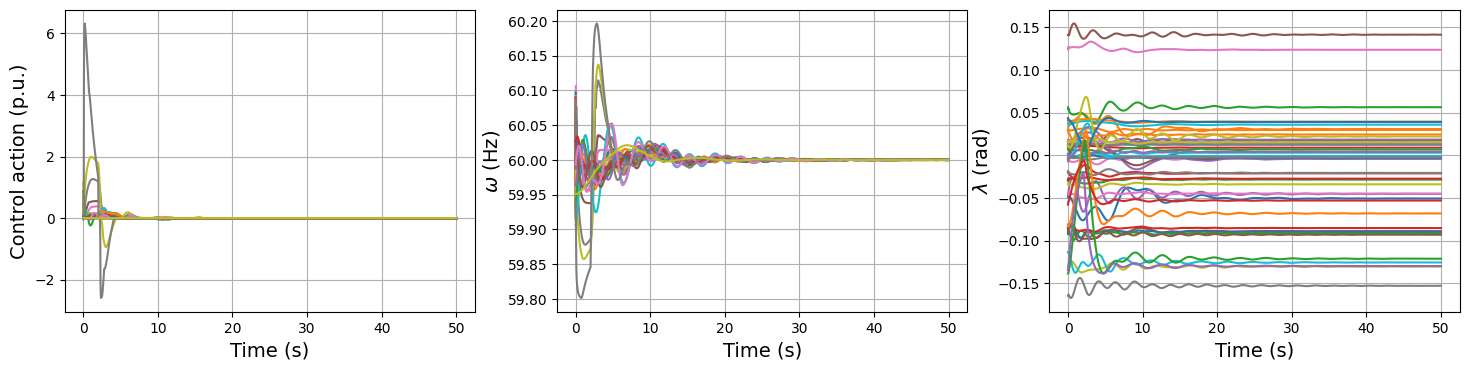}}\\
    \caption{Dynamics of the IEEE 39-bus network under the trained controllers via the RLb method when only partial information about the sudden change on power injection is known in the training process. In this case, 
    any of the buses 32, 33, 35 and 38 may encounter a sudden change on power injection.
 }\label{fig:sim-add}
\end{figure*}

\section{Case Study}
We conduct a case study to illustrate the performance of the RLb method. We consider the IEEE 39-bus power network and assume each bus represents an aggregate area containing loads and generators with system parameters from~\cite{YZ-JC:19-auto,KWC-JC-GR:09}. We consider the time horizon of interest to be $T=50$ seconds and bus 38 encountering a sudden change in power injection during the time interval $(0,2]$ seconds, with $\Delta p_{38} = -p^{\text{nom}}_{38}$. The nominal frequency is 60 Hz, and the safe region is set as [59.8, 60.2] Hz for every bus.


\subsection{Comparison baseline and training setup}
We conduct the following comparison for illustration. The first is to compare RLb method to the methods proposed in~\cite{WC-YJ-BZ:22} and~\cite{YZ-JC:19-auto}. The controllers in~\cite{WC-YJ-BZ:22} are parameterized as non-increasing functions passing through the origin without saturation limits and then trained via RL, and the controllers in~\cite{YZ-JC:19-auto}
are designed as $u_{i}(x, p) = \min \{0, \frac{\overline{\alpha}_{i}(\overline{\omega}_{i}-\omega_{i})}{\omega_{i}-\overline{\omega}_{i}\threshold}+q_{i}(x, p)\}$ for $\omega_{i}>\overline{\omega}_{i}\threshold$, $u_{i}(x, p) = \max \{0, \frac{\underline{\alpha}_{i}(\underline{\omega}_{i}-\omega_{i})}{\underline{\omega}_{i}\threshold-\omega_{i}}+q_{i}(x, p)\}$ for $\omega_{i}<\underline{\omega}_{i}\threshold$, and $u_{i}(x, p) = 0$  otherwise, with $\overline{\alpha}_i(s) = \underline{\alpha}_i(s) = 2s$ and $\overline{\omega}\threshold_i = 0.1$, $\underline{\omega}\threshold_i = -0.1$ for all $i \in \Ic$.

The RL environment is built using TensorFlow 2.7.0 and the training process is conducted in Google Colab on a single TPU with 32 GB memory. 
We use the same training hyper parameters for both the RL-based method in~\cite{WC-YJ-BZ:22} and the RLb method to solve~\eqref{eq:NN-optimization}.
The discretization stepsize $\Delta t$ is set as 0.0008 seconds. To facilitate the training process, we only evaluate the first 4 seconds in each episode, meaning the total number of stages $K$ in each episode is 5000. For each $i \in \Ic$, $\omega_i(0)$ is randomly generated in [59.9, 60.1] Hz, and $\lambda(0)$ is calculated using power injections randomly generated over [0.9$p_i^{\text{nom}}(0)$, 1.1$p_i^{\text{nom}}(0)$]. The balancing coefficient in the objective function is $\gamma=40$, and the number of episodes $I$, the batch size $H$, and the number of neurons $m$ are 200, 50, 20, respectively. We use the Adam algorithm \cite{DPK-JB:15} to update the parameter $\phi$ in each episode with learning rate $\alpha=0.2$.
 

\subsection{Simulation results}

\begin{figure}[ht]
    \centering
    \includegraphics[width=.75\linewidth]{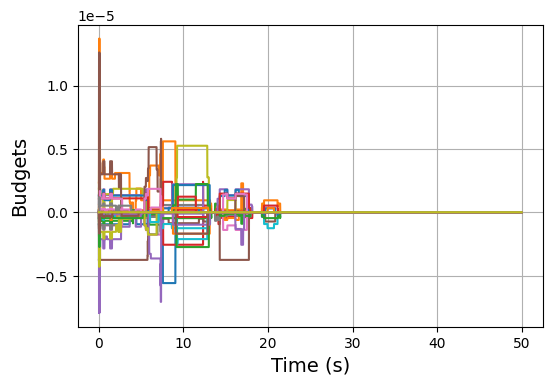}
    \caption{Budget allocation under the proposed dynamic budget assignment mechanism for the RLb method.}
    \label{fig:budgets}
\end{figure}

We fist consider the case that we know exactly which bus encounters the sudden change on power injection in the training process.
Fig.~\ref{fig:loss} compares the training loss curves of the RL-based method in~\cite{WC-YJ-BZ:22} and the RLb method.
Fig.~\ref{fig:sim} illustrates the performance of the RLb method, the RL-based method in~\cite{WC-YJ-BZ:22}, and the method in~\cite{YZ-JC:19-auto} with the same randomly generated initial states.  
Table~\ref{tab:simu-comparison} summarizes the comparison results. The RLb method guarantees both asymptotic stability
and transient safety, while significantly reducing the cost compared to the method in~\cite{YZ-JC:19-auto}.
The RLb method also achieves faster convergence and smaller transient fluctuations.
Note that the RLb method leads to a greater cost compared to the one in~\cite{WC-YJ-BZ:22}, which is reasonable since it uses a greater control effort to guarantee the frequency safety during the transient, cf. Fig.~\ref{fig:sim}. Fig.~\ref{fig:budgets} shows the dynamic budget allocation under the proposed assignment mechanism, validating that the budget for each bus vanishes when its frequency converges while keeping the summation of all the budgets equal to zero. In this way, more nodes can contribute to the transient frequency regulation while cooperatively minimizing the cost. 

We next test the case when there is only partial information about the sudden change on power injection. In the training process, we assume that any of the buses 32, 33, 35 and 38 may encounter a sudden change on power injection. Fig.~\ref{fig:sim-add} illustrates the performance of the RLb method in this case under the same initial states as in Fig.~\ref{fig:sim}.  Table~\ref{tab:simu-comparison} reports the cost, showing that the RLb method remains effective even without exact information about the sudden change on power injection.

\begin{figure*}[htb]
  \centering
{
  \includegraphics[width=.9\linewidth]{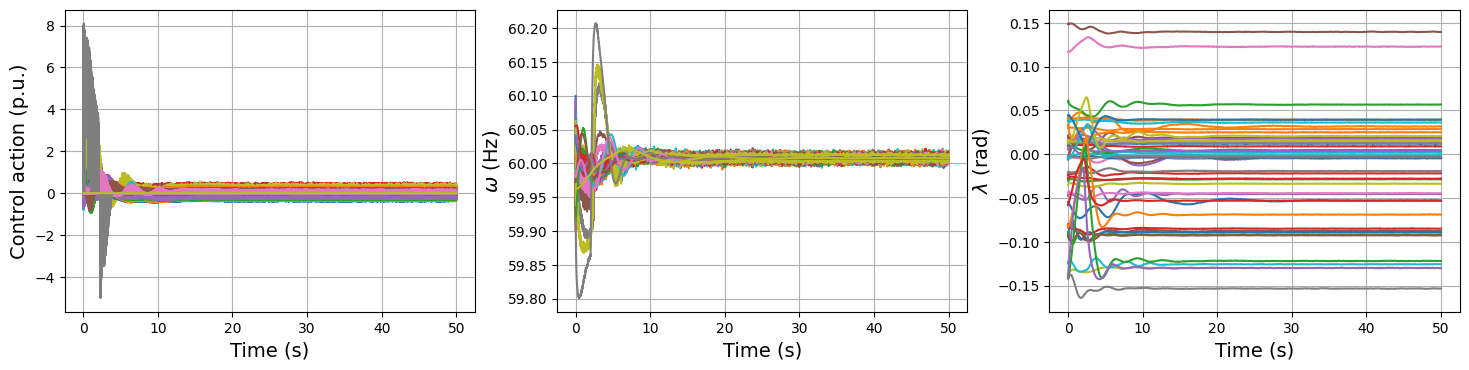}}
  \\
   {\includegraphics[width=.9\linewidth]{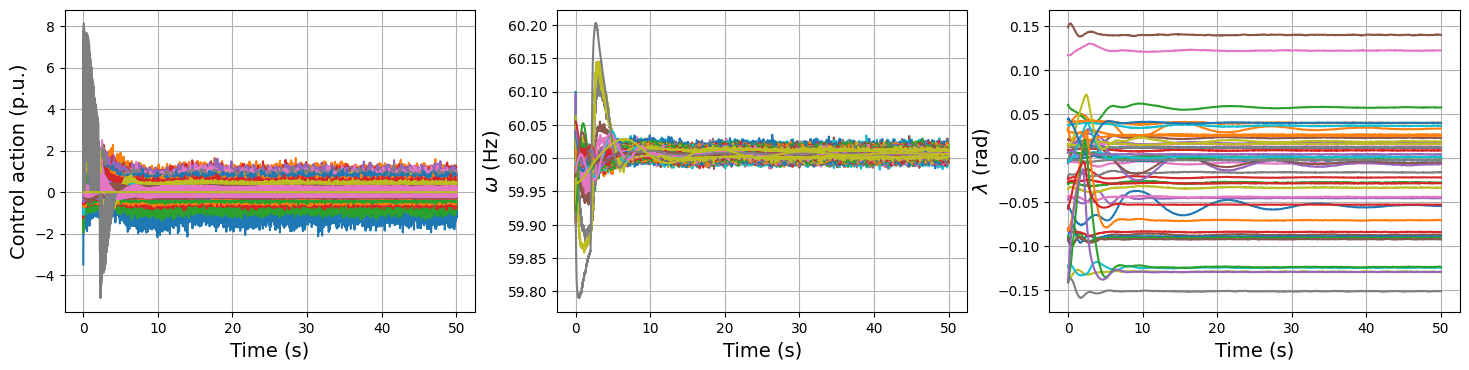}}
    \caption{Dynamics of the IEEE 39-bus network under the proposed RL-based controller with (top) and  without (bottom) Lipschitz regularization. The frequency measurement noise is uniformly randomly generated over $[-0.05,0.05]$ Hz. The parameter $\rho$ is set as 500.
 }\label{fig:sim-robust}
\end{figure*}

Finally, Fig.~\ref{fig:sim-robust} compares the performance of the RLb method with and without Lipschitz regularization in the presence of frequency measurement noise. In both cases the frequency nadir slightly violates the safety bound due to the noisy input to controllers. However, the Lipschitz regularization helps enhance system robustness by reducing the fluctuations in state and control.

\begin{table}[htb]
\caption{Comparison Results\protect\footnotemark}
    \label{tab:simu-comparison}
    \centering
        \begin{tabular}{c||c|c|c}\hline
        \hline
            Method & Stability & Safety  &Cost \\ \hline
            \cite{WC-YJ-BZ:22} & \checkmark  & $\times$ & 0.9510 \\ \hline
            \cite{YZ-JC:19-auto} & \checkmark  & \checkmark & 2.2054 \\ \hline
            RLb & \checkmark  & \checkmark & 1.1628 \\ \hline
            RLb* & \checkmark  & \checkmark & 1.1721 \\ \hline
        \end{tabular}\\
\end{table}

\footnotetext{RLb* denotes the RLb method with only partial information about where the sudden change on power injection might occur.}

\section{Conclusions}
We have presented a reinforcement learning approach to the synthesis of optimal controllers that are distributed and guarantee the stability and transient safety of power networks. Leveraging notions of Lyapunov stability and safety-critical control, we have identified conditions on the controller design that ensure stability and transient frequency safety. These constraints incorporate the idea of endowing some buses with additional design flexibility through budgets in a way that collectively ensures the stability of the overall system. We have constructed neural networks to parameterize the control policies within the identified search space and employed an RL framework to learn an optimal controller. Simulations illustrate the guaranteed stability and transient frequency safety of the resulting closed-loop system while showing a significant reduction in  the cost. Our future work aims to refine dynamic budget allocation policies to further reduce cost and 
transient fluctuation, e.g., by devising state-dependent mechanisms that allow agents to negotiate exchanges in their budgets, consider feasibility constraints on the control input, and incorporate higher-order dynamics of generators, inverter-interfaced energy resources, and their existing control loops.

\vspace{-2ex}
\section*{CRediT authorship contribution statement}
\textbf{Z. Yuan:} Formal Analysis, Writing - Original Draft, Visualization. \textbf{C. Zhao:} Writing - Review \& Editing, Supervision. \textbf{J. Cort\'es:} Conceptualization, Formal Analysis, Writing - Review \& Editing, Supervision.

\vspace{-2ex}
\section*{Declaration of competing interest}
The authors declare that they have no known competing financial interests or personal relationships that could have appeared
to influence the work reported in this paper.

\vspace{-2ex}
\section*{Acknowledgments}
The work of Z. Yuan and J. Cort\'es was supported by NSF Award 2044900, and the work of C. Zhao was supported by Hong Kong Research Grants Council through General Research Fund 14212822.

\biboptions{sort&compress}
\bibliographystyle{unsrt}

\end{sloppypar}
\end{document}